\pdfoutput=1

\documentclass[11pt,dvipsnames]{article}

\usepackage[]{naacl2021}

\usepackage{times}
\usepackage{latexsym}
\usepackage{subcaption}
\usepackage{adjustbox}
\usepackage{multirow}

\usepackage[T1]{fontenc}

\usepackage[utf8]{inputenc}

\usepackage{microtype}
\usepackage{amsmath}
\usepackage{amssymb}
\usepackage{bm}
\usepackage{graphicx}
\usepackage{xcolor}

\title{COIL: Revisit Exact Lexical Match in Information Retrieval \\ with Contextualized Inverted List }

\author{Luyu Gao,   Zhuyun Dai, Jamie Callan \\
  Language Technologies Institute \\
  Carnegie Mellon University \\
  \texttt{\{luyug, zhuyund, callan\}@cs.cmu.edu}  }

\begin{document}
\maketitle
\begin{abstract}
Classical information retrieval systems such as BM25 rely on exact lexical match and carry out search efficiently with inverted list index. Recent neural IR models shifts towards soft semantic matching all query document terms, but they lose the computation efficiency of exact match systems. 
This paper presents COIL, a contextualized exact match retrieval architecture that brings semantic lexical matching. COIL scoring is based on overlapping query document tokens' contextualized representations. The new architecture stores contextualized token representations in inverted lists,  bringing together the efficiency of exact match and the representation power of deep language models. Our experimental results show COIL outperforms classical lexical retrievers and state-of-the-art deep LM retrievers with similar or smaller latency.\footnote{Our code is available at \url{https://github.com/luyug/COIL}.}

\end{abstract}
\section{Introduction}
Widely used, bag-of-words~(BOW) information retrieval~(IR) systems such as BM25 rely on exact lexical match~\footnote{Exact match up to morphological changes.} between query and document terms. Recent study in neural IR takes a different approach and compute soft matching between all query and document terms to model complex matching. 


The shift to soft matching in neural IR models attempts to address \emph{vocabulary mismatch} problems, that query and the relevant documents use different terms, e.g. cat v.s. kitty, for the same concept~\cite{dssm,DRMM,KNRM}.
Later introduction of contextualized representations~\cite{ELMo} from deep language models~(LM) further address \emph{semantic mismatch}, that the same term can refer to different concepts, e.g., bank of river vs. bank in finance. 
Fine-tuned deep LM rerankers produce token representations based on context and achieve state-of-the-art in text ranking with huge performance leap \cite{Nogueira2019PassageRW, DBLP:conf/sigir/DaiC19}.

Though the idea of soft matching all tokens is carried through the development of neural IR models, seeing the success brought by deep LMs, we take a step back and ask: how much gain can we get if we introduce contextualized representations back to lexical exact match systems? In other words, can we build a system that still performs exact query-document token matching but compute matching signals with contextualized token representations instead of heuristics? 
This may seem a constraint on the model, but exact lexical match produce more explainable and controlled patterns than soft matching. It also allows search to focus on only the subset of documents that have overlapping terms with query, which can be done efficiently with inverted list index. Meanwhile, using dense contextualized token representations enables the model to handle semantic mismatch,  which has been a long-standing problem in classic lexical systems.

To answer the question, we propose a new lexical matching scheme that uses vector similarities between query-document overlapping term contextualized representations to replace heuristic scoring used in classical systems. We present COntextualized Inverted List~(COIL), a new exact lexical match retrieval architecture armed with deep LM representations.  
COIL processes documents with deep LM offline and produces representations for each document token. 
The representations are grouped by their surface tokens into inverted lists. At search time, we build representation vectors for query tokens and perform contextualized exact match: use each query token to look up \emph{its own} inverted list and compute vector similarity with document vectors stored in the inverted list as matching scores. COIL enables efficient search with rich-in-semantic matching between query and document. 

Our contributions include 1) introduce a novel retrieval architecture, \textit{contextualized inverted lists}~(COIL) that brings semantic matching into lexical IR systems, 2) show matching signals induced from exact lexical match can capture complicated matching patterns, 3) demonstrate COIL significantly outperform classical and deep LM augmented lexical retrievers as well as state-of-the-art dense retrievers on two retrieval tasks. 
\begin{figure*}[t]
\centering
    \begin{subfigure}[t]{0.46\textwidth}
        \centering
        \includegraphics[width=0.96\textwidth]{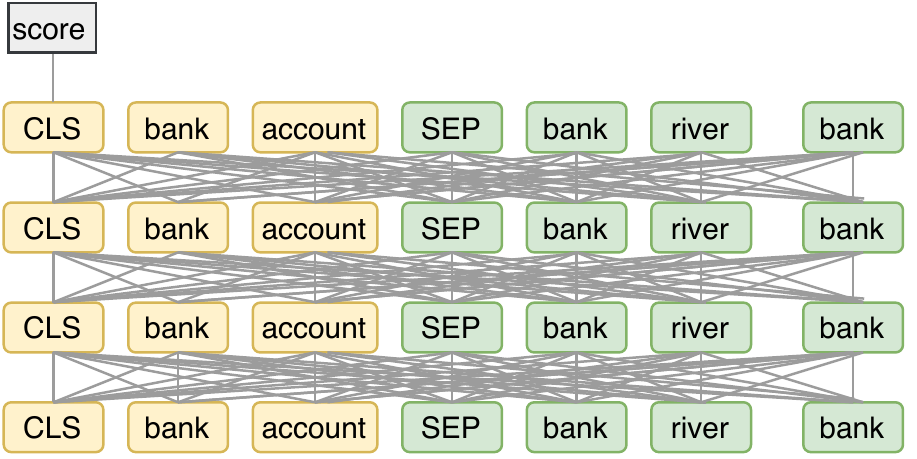}
        \caption{Cross-Attention Model (e.g., BERT reranker)}
        \label{fig:cross-attn-model}
    \end{subfigure}~
\begin{subfigure}[t]{0.46\textwidth}
        \centering
        \includegraphics[width=0.96\textwidth]{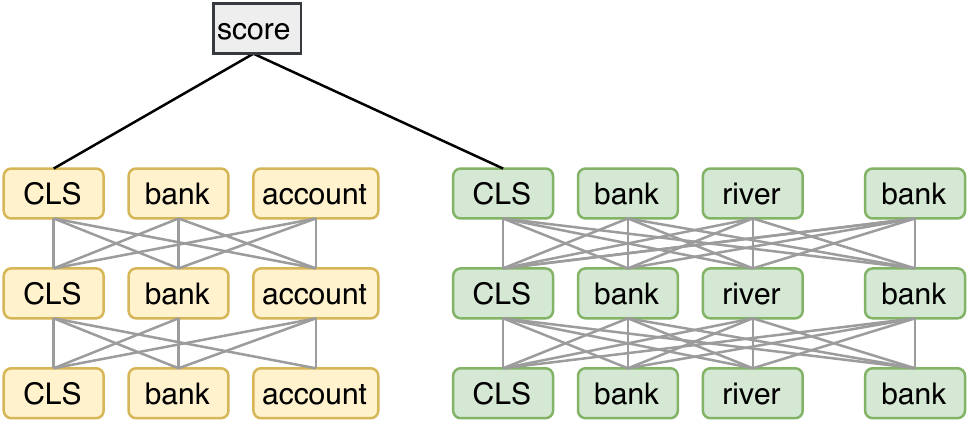}
        \caption{Dense Retrievers (e.g., DPR)}
        \label{fig:dense-model}
    \end{subfigure}%
    
    \begin{subfigure}[t]{0.46\textwidth}
        \centering
        \includegraphics[width=0.98\textwidth]{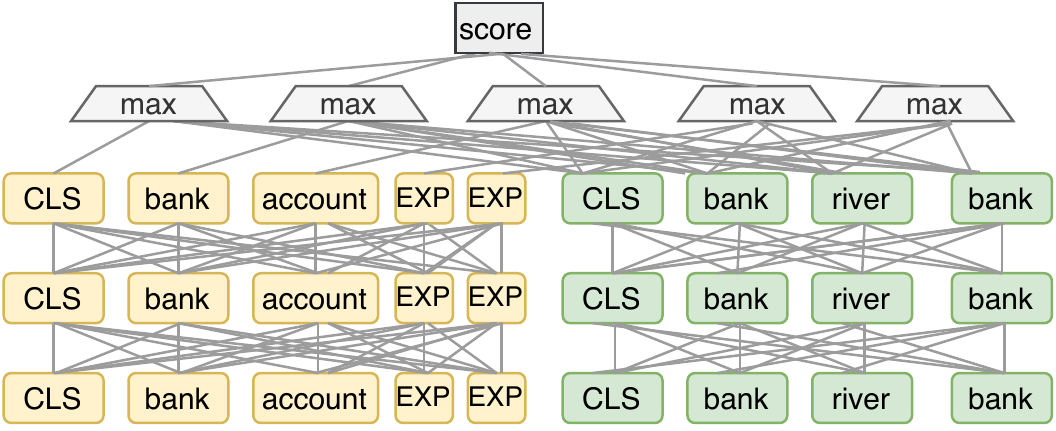}
        \caption{ ColBERT: All-to-All Match}
        \label{fig:late-model}
    \end{subfigure}%
~
        \begin{subfigure}[t]{0.46\textwidth}
        \centering
        \includegraphics[width=0.96\textwidth]{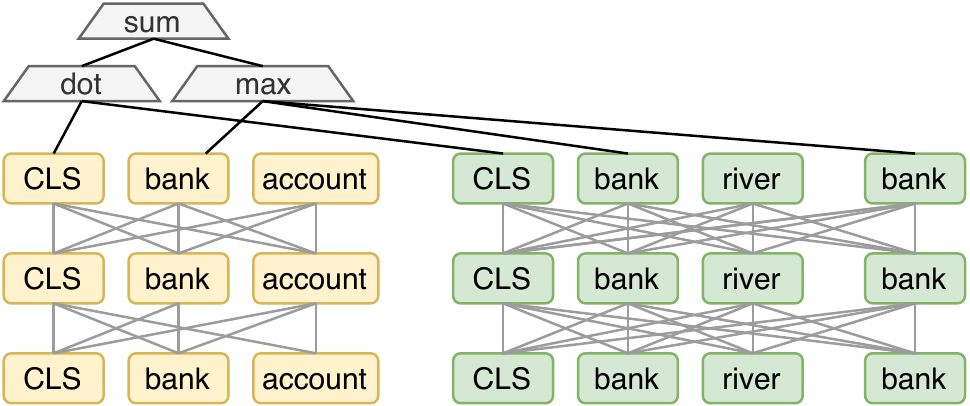}
        \caption{COIL: Contextualized Exact Match }\label{fig:coil-model}
    \end{subfigure}
\caption{An illustration of reranking/retrieval mechanisms with deep LM, including our proposed model, COIL.}
\label{fig: coil}
\end{figure*}

\section{Related Work}

\paragraph{Lexical Retriever} Classical IR systems rely on exact lexical match retrievers such as Boolean Retrieval, BM25~\cite{bm25} and statistical language models~\cite{lafferty2001document}. 
This type of retrieval model can process
queries very quickly by organizing the documents into inverted index, where each distinct term has an inverted list that 
stores information about documents it appears in. Nowadays, they are still widely used in production systems. 
However, these retrieval models fall short of matching related terms (vocabulary mismatch) or modeling context of the terms (semantic mismatch). Much early effort was put into improving exact lexical match retrievers, such as matching n-grams~\cite{metzler2005markov} or expanding queries with terms from related documents~\cite{Lavrenko2001RelevanceBasedLM}. However, these methods still use BOW framework and have limited capability of modeling human languages. 

\paragraph{Neural Ranker} In order to deal with vocabulary mismatch, neural retrievers that rely on soft matching between numerical text representations are introduced. Early attempts compute similarity between pre-trained word embedding such as word2vec~\cite{w2v} and GLoVe~\cite{glove} to produce matching score~\cite{ganguly2015word, diazquery}. One more recent approach encodes query and document each into a vector and computes vector similarity~\cite{dssm}. Later researches realized the limited capacity of a single vector to encode fine-grained information and introduced full interaction models to perform soft matching between all term vectors~\cite{DRMM,KNRM}. In these approaches, scoring is based on learned neural networks and the hugely increased computation cost limited their use to reranking a top candidate list generated by a lexical retriever.

\paragraph{Deep LM Based Ranker and Retriever} Deep LM made a huge impact on neural IR. Fine-tuned Transformer~\cite{transformer} LM BERT~\cite{BERT} achieved state-of-the-art reranking performance for passages and documents~\cite{Nogueira2019PassageRW, DBLP:conf/sigir/DaiC19}. As illustrated in \autoref{fig:cross-attn-model}, the common approach is to feed the concatenated query document text through BERT
and use BERT’s [CLS] output token to produce a
relevance score.
The deep LM rerankers addressed both vocabulary and semantic mismatch by computing full cross attention between contextualized token representations. 
Lighter deep LM rankers are developed~\cite{MacAvaney2020EfficientDR,MORES}, but their cross attention operations are still too expensive for full-collection retrieval.

Later research therefore resorted to augmenting lexical retrieval with deep LMs by expanding the document surface form to narrow the vocabulary gap, e.g., DocT5Query~\cite{docTTTTTquery}, or altering term weights to emphasize important terms, e.g., DeepCT~\cite{DeepCT}. Smartly combining deep LM retriever and reranker can offer additive gain for end performance~\cite{LCE}. These retrievers however still suffer from vocabulary and semantic mismatch as traditional lexical retrievers. 

Another line of research continues the work on single vector representation and build dense retrievers, as illustrated in \autoref{fig:dense-model}. They store document vectors in a dense index and retrieve them through Nearest Neighbours search. Using deep LMs, dense retrievers have achieved promising results on several retrieval tasks~\cite{DPR}. Later researches show that dense retrieval systems can be further improved by better training~\cite{ANCE, CLEAR}.

Single vector systems have also been extended to multi-vector representation systems. Poly-encoder~\cite{Humeau2020PolyencodersAA} encodes queries into a set of vectors. Similarly, Me-BERT~\cite{Luan2020SparseDA} represents documents with a set of vectors. A concurrent work ColBERT (\autoref{fig:late-model}) use multiple vectors to encode both queries and documents~\cite{ColBERT}. In particular, it represents a documents with all its terms' vectors and a query with an expanded set of term vectors. It then computes all-to-all~(Cartesian) soft match between the tokens. ColBERT performs interaction as dot product followed pooling operations, which allows it to also leverage a dense index to do full corpus retrieval. However,  since ColBERT encodes a document with all tokens, it adds another order of magnitude of index complexity to all aforementioned methods: document tokens in the collection need to be stored in a \emph{single} huge index and considered at query time. Consequently, ColBERT is engineering and hardware demanding.


\section{Methodologies}
\label{sec:method}
In this section, we first provide some preliminaries on exact lexical match systems. Then we discuss COIL's contextualized exact match design and how its search index is organized. We also give a comparison between COIL and other popular retrievers.
\begin{figure}[t]
    \centering
    \includegraphics[width=0.49\textwidth]{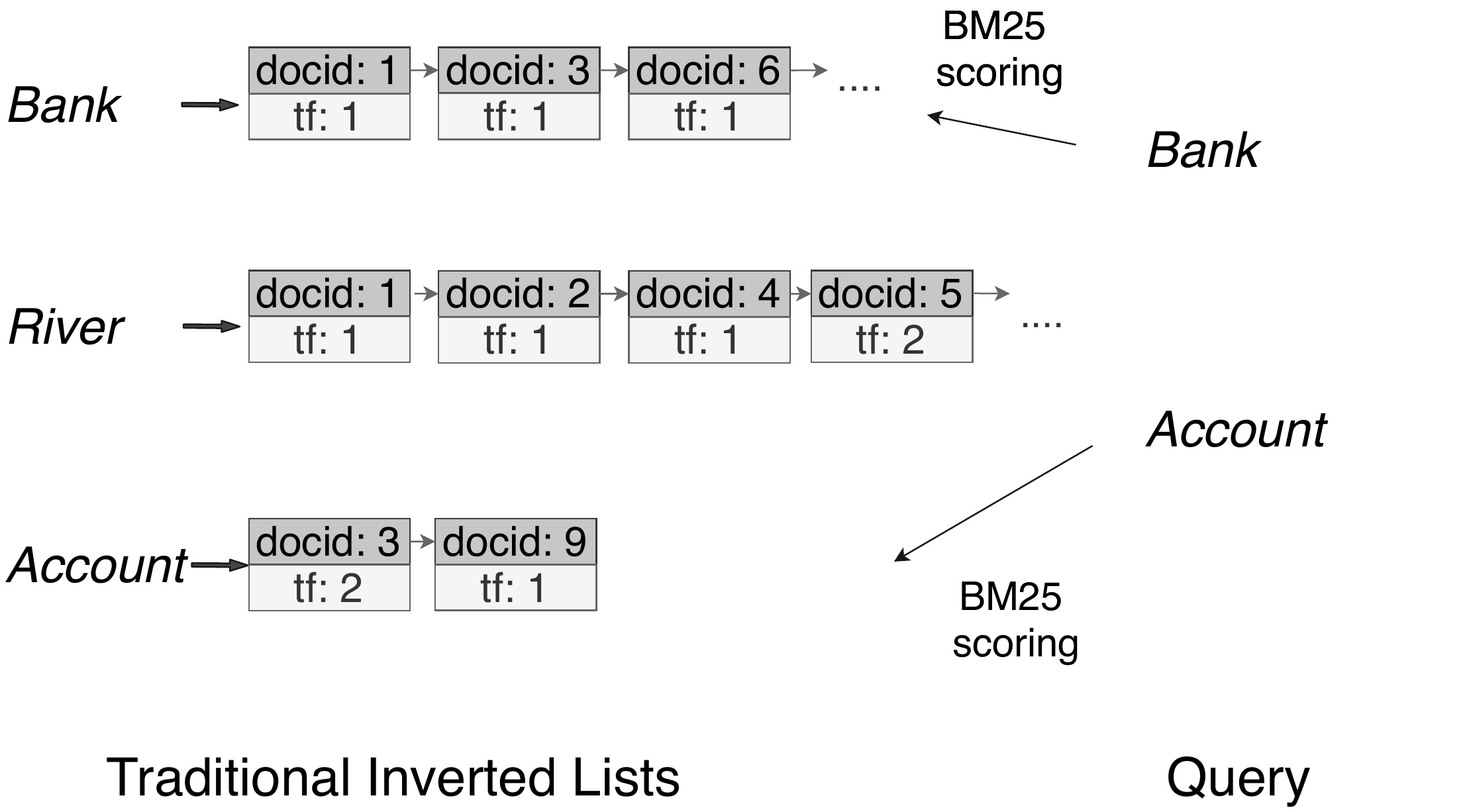}
    \caption{An illustration of traditional inverted lists. The inverted list maps a term to the list of documents where the term occurs. Retriever looks up query terms' inverted lists and scores those documents with stored statistics such as term frequency (tf).}
    \label{fig:traditional-index}
\end{figure}

\begin{figure}[t]
    \centering
    \includegraphics[width=0.47\textwidth]{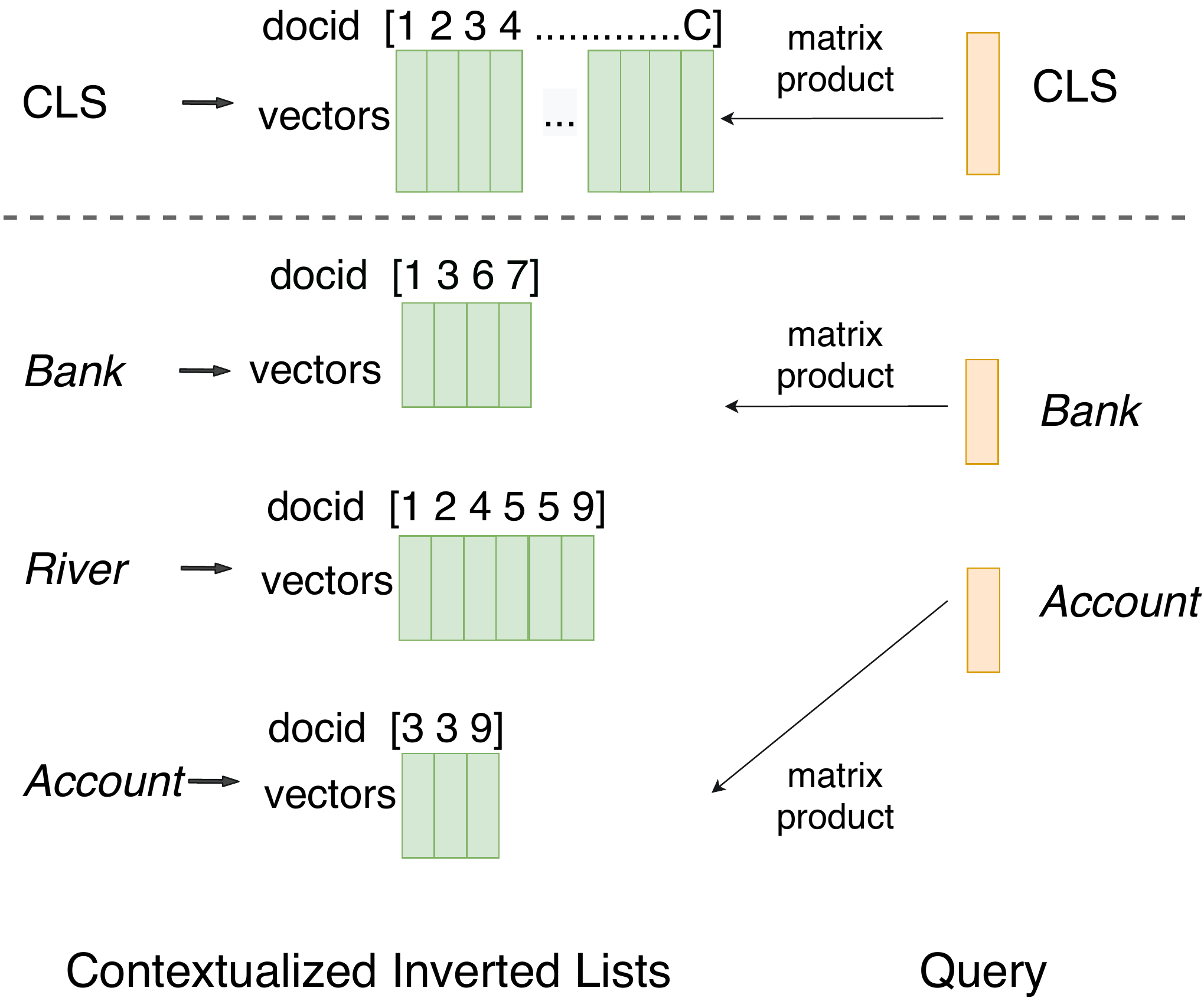}
    \caption{COIL's index and retrieval architecture.  COIL-tok relies on the exact token matching (lower). COIL-full includes in addition CLS matching (upper).}
    \label{fig: coil-index}
\end{figure}

\subsection{Preliminaries}
Classic lexical retrieval system relies on \emph{overlapping} query document terms under morphological generalization like stemming, in other words, \emph{exact lexical match}, to score query document pair. A scoring function is defined as a sum of matched term scores. The scores are usually based on statistics like term frequency (\textit{tf}). Generally, we can write,
\begin{equation}
    \label{eq:general}
    s = \sum_{t \in q \cap d} \sigma_t (h_q(q, t), h_d(d, t))
\end{equation}
where for each overlapping term $t$ between query $q$ and document $d$, functions $h_q$ and $h_d$ extract term information and a term scoring function $\sigma_t$ combines them.
A popular example is BM25, which computes,
\begin{equation}
\begin{split}
    &s_{\text{BM25}} = \sum_{t \in q \cap d} idf(t) h_q^\text{BM25}(q, t) h_d^\text{BM25}(d, t) \\
    &h_q^\text{BM25}(q, t) = \frac{tf_{t,q}(1+k_2)}{tf_{t,q} + k_2} \\
    &h_d^\text{BM25}(d, t) = \frac{tf_{t,d}(1+k_1)}{tf_{t,d} + k_1 (1-b + b \frac{|d|}{\text{avgdl}})} 
\end{split}
\label{eq:bm25}
\end{equation}
where $tf_{t, d}$ refers to term frequency of term $t$ in document $d$, $tf_{t,q}$ refers to the term frequency in query, $idf(t)$ is inverse document frequency, and $b$, $k_1$, $k_2$ are hyper-parameters. 




One key advantage of exact lexical match systems lies in efficiency. With summation over exact matches, scoring of each query term only goes to documents that contain matching terms.
This can be done efficiently using \emph{inverted list} indexing (\autoref{fig:traditional-index}). The inverted list maps back from a term to a list of documents where the term occurs. To compute \autoref{eq:general}, the retriever only needs to traverse the subset of documents in query terms' inverted lists instead of going over the entire document collection.

While recent neural IR research mainly focuses on breaking the exact match bottleneck with soft matching of text, we hypothesize that exact match itself can be improved by replacing semantic independent frequency-based scoring with semantic rich scoring.
In the rest of this section, we show how to modify the exact lexical match framework with contextualized term representations to build effective and efficient retrieval systems. 

\subsection{Contextualized Exact Lexical Match}
\label{sec:arch}
Instead of term frequency, we desire to encode the semantics of terms to facilitate more effective matching. Inspired by recent advancements in deep LM, we encode \emph{both} query and document tokens into contextualized vector representations and carry out matching between exact lexical matched tokens. \autoref{fig:coil-model} illustrates the scoring model of COIL. 

In this work, we use a Transformer language model\footnote{We used the base, uncased variant of BERT.}  as the contextualization function.
We encode a query $q$ with the language model~(LM) and represent its $i$-th token by projecting the corresponding output: 
\begin{equation}
\begin{aligned}
    \bm{v}_i^q &= \bm{W}_{tok} \text{LM}(q, i) + \bm{b}_{tok}
    \label{eq:proj-q-tok}
\end{aligned}
\end{equation}
where $\bm{W}_{tok}^{n_t \times n_{lm}}$ is a matrix that maps the LM's $n_{lm}$ dimension output into a vector of lower dimension $n_t$. We down project the vectors as we hypothesize that it suffices to use lower dimension token vectors. We confirm this in \autoref{sec:results}. 
Similarly, we encode a document $d$'s $j$-th token $d_j$ with:
\begin{equation}
\begin{aligned}
    \bm{v}_j^d &= \bm{W}_{tok} \text{LM}(d, j) + \bm{b}_{tok}
    \label{eq:proj-d-tok}
\end{aligned}
\end{equation}

We then define the \emph{contextualized exact lexical match scoring function} between query document based on vector similarities between exact matched query document token pairs:
\begin{equation}
    s_\text{tok}(q, d) 
    = \sum_{q_i \in q \cap d} \underset{d_j = q_i}{\max} ({\bm{v}_i^q}^\intercal \bm{v}^d_j)
    \label{eq:score-max}
\end{equation}
Note that, importantly, the summation goes through only overlapping terms, $q_i \in q \cap d$. For each query token $q_i$, we finds all \emph{same} tokens $d_j$ in the document, computes their similarity with $q_i$ using the \emph{contextualized} token vectors. The maximum similarities are picked for query token $q_i$. Max operator is adopted to capture the most important signal~\cite{textCNN}. This fits in the general lexical match formulation, with $h_q$ giving representation for $q_i$, $h_t$ giving representations for all $d_j=q_i$, and $\sigma_t$ compute dot similarities between query vector with document vectors and max pool the scores.

As with classic lexical systems, $s_{tok}$ defined in \autoref{eq:score-max} does not take into account similarities between lexical-different terms, thus faces vocabulary mismatch. 
Many popular LMs~\cite{BERT,XLNet,RoBERTa} use a special CLS token to aggregate sequence representation. We project the CLS vectos with $\bm{W}_{cls}^{n_c \times n_{lm}}$ to represent the entire query or document, 
\begin{equation}
\begin{aligned}
    \bm{v}_{cls}^q &= \bm{W}_{cls} \text{LM}(q, \text{CLS}) + \bm{b}_{cls} \\
    \bm{v}_{cls}^d &= \bm{W}_{cls} \text{LM}(d, \text{CLS}) + \bm{b}_{cls}
    \label{eq:proj-cls}
\end{aligned}
\end{equation}
The similarity between $\bm{v}_{cls}^q$ and $\bm{v}_{cls}^d$ provides high-level semantic matching and mitigates the issue of vocabulary mismatch. The full form of COIL is:
\begin{equation}
    s_\text{full}(q, d) 
    = s_\text{tok}(q, d)  + {\bm{v}_{cls}^q}^\intercal \bm{v}_{cls}^d 
\end{equation}
In the rest of the paper, we refer to systems with CLS matching \textbf{COIL-full} and without \textbf{COIL-tok}.

COIL's scoring model (\autoref{fig:coil-model}) is fully differentiable. Following earlier work~\cite{DPR}, we train COIL with negative log likelihood defined over query $q$, a positive document $d^+$ and a set of negative documents $\{d^-_1, d^-_2, .. d^-_l ..\}$ as loss. 
\begin{equation}
    \mathcal{L} = -\log \frac{\exp(s(q, d^+))}{ \exp(s(q, d^+)) + \underset{l}{\sum} \exp(s(q, d^-_l)) }
\end{equation}
Following \citet{DPR}, we use in batch negatives and hard negatives generated by BM25. Details are discussed in implementation, \autoref{sec: exp-method}.

\subsection{Index and Retrieval with COIL}
\label{sec:retrive-and-index}
COIL pre-computes the document representations and builds up a search index, which is illustrated in \autoref{fig: coil-index}. Documents in the collection are encoded offline into token and CLS vectors. Formally, for a unique token $t$ in the vocabulary $V$, we collect its contextualized vectors from all of its mentions from documents in collection $C$, building token $t$'s contextualized inverted list:
\begin{equation}
    I^t = \{ \bm{v}_j^{d} \; | \; d_j=t , d \in C \},
\end{equation}
where $\bm{v}_j^{d}$ is the BERT-based token encoding defined in \autoref{eq:proj-d-tok}. We define search index to store inverted lists for all tokens in vocabulary, $\mathbb{I}=\{ I^t \; | \; t \in V\}$. For COIL-full, we also build an index for the CLS token $I^{cls} = \{ \bm{v}_{cls}^{d} \; | \; d \in C \}$ . 

As shown in \autoref{fig: coil-index}, in this work we implement COIL's  by stacking vectors in each inverted list $I^t$ into a matrix $M^{n_t \times |I^k|}$, 
so that similarity computation that traverses an inverted list and computes vector dot product can be done efficiently as one matrix-vector product with optimized BLAS~\cite{blas} routines on CPU or GPU. 
All $\bm{v}_{cls}^d$ vectors can also be organized in a similar fashion into matrix $M_{cls}$ and queried with matrix product. The matrix implementation here is an exhaustive approach that involves all vectors in an inverted list.
As a collection of dense vectors, it is also possible to organize each inverted list as an approximate search index~\cite{faiss, scann} to further speed up search. 

When a query $q$ comes in, we encode every of its token into vectors $\bm{v}_i^q$. The vectors are sent to the subset of COIL inverted lists that corresponds query tokens $\mathbb{J}=\{ I^t \; | \; t \in q\}$. where the matrix product described above is carried out. This is efficient as $|\mathbb{J}| << |\mathbb{I}|$, having only a small subset of all inverted lists to be involved in search. 
For COIL-full, we also use encoded CLS vectors $\bm{v}_{cls}^q$ to query the CLS index to get the CLS matching scores. The scoring over different inverted lists can serve in parallel. The scores are then combined by \autoref{eq:score-max} to rank the documents.

Readers can find detailed illustration figures in the \autoref{sec:appendix}, for index building and querying, \autoref{fig:index-build} and \autoref{fig:index-search}, respectively.

\subsection{Connection to Other Retrievers}
\paragraph{Deep LM based Lexical Index} 
Models like DeepCT~\cite{DeepCT, HDCT} and DocT5Query~\cite{docTTTTTquery} alter $tf_{t,d}$ in documents with deep LM BERT or T5. This is similar to a COIL-tok with token dimension $n_t = 1$. A single degree of freedom however measures more of a term \emph{importance} than \emph{semantic agreement}. 

\paragraph{Dense Retriever}  Dense retrievers (\autoref{fig:dense-model}) are equivalent to COIL-full's CLS matching. COIL makes up for the lost token-level interactions in dense retriever with exact matching signals.

\paragraph{ColBERT} ColBERT (\autoref{fig:late-model}) computes relevance by soft matching \emph{all} query and document term's contextualized vectors. 
\begin{equation}
    s(q,d) = \sum_{q_i \in [cls;q;exp]} \max_{d_j \in [cls;d]} ({\bm{v}_i^q}^\intercal \bm{v}^d_j)
\end{equation}
where interactions happen among query $q$, document $d$, $cls$ and set of query expansion tokens $exp$. The all-to-all match contrasts COIL that only uses exact match. 
It requires a dense retrieval over all document tokens' representations as opposed to COIL which only considers query's overlapping tokens, and are therefore much more computationally expensive than COIL. 


\section{Experiment Methodologies}
\label{sec: exp-method}


\paragraph{Datasets}
We experiment with two large scale ad hoc retrieval benchmarks from the TREC 2019 Deep Learning (DL) shared task: MSMARCO passage~(8M English passages of average length around 60 tokens) and MSMARCO document~(3M English documents of average length around 900 tokens)\footnote{Both datasets can be downloaded from \url{https://microsoft.github.io/msmarco/}}. For each, we train models with the MSMARCO Train queries, and record results on MSMARCO Dev queries and TREC DL 2019 test queries.  We report mainly full-corpus retrieval results but also include the rerank task on MSMARCO Dev queries where we use neural scores to reorder BM25 retrieval results provided by MSMARO organizers. Official metrics include MRR@1K and NDCG@10 on test and MRR@10 on MSMARCO Dev. We also report recall for the dev queries following prior work~\cite{DeepCT, docTTTTTquery}. 

\begin{table*}[t]
\caption{MSMARCO passage collection results. Results not applicable are denoted `--' and no available `n.a.'.}
\centering
\renewcommand{\arraystretch}{0.9}
\begin{tabular}{ l || c | c c | c c}
\hline \hline
 & \multicolumn{4}{c}{MS MARCO Passage Ranking}\\ 
\hline
 & Dev Rerank & \multicolumn{2}{c|}{Dev Retrieval} & \multicolumn{2}{c}{ DL2019 Retrieval} \\ 
 
Model & MRR@10 & MRR@10 & Recall@1K & NDCG@10 & MRR@1K  \\  
\hline
\textbf{Lexical Retriever} & & & & & \\
BM25  & -- &  0.184	& 0.853 & 0.506 & 0.825 \\
DeepCT & -- & 0.243 & 0.909 & 0.572  & 0.883 \\
DocT5Query & -- & 0.278 & 0.945 & 0.642 & 0.888  \\
BM25+BERT reranker  & 0.347 &  --	& -- & -- &  --  \\

\hline
\textbf{Dense Retriever} & & & & & \\

Dense (BM25 neg) & n.a. & 0.299 & 0.928 & 0.600 & n.a.\\
Dense (rand + BM25 neg) & n.a.  & 0.311 & 0.952 & 0.576 & n.a.\\
Dense (our train) & 0.312 & 0.304 & 0.932 & 0.635 & 0.898  \\
\hline
ColBERT & 0.349 &  0.360 & 0.968 & n.a. & n.a.\\
\hline
COIL-tok & 0.336 & 0.341 & 0.949 & 0.660 & 0.915 \\
COIL-full & 0.348 & 0.355 & 0.963 & 0.704 & 0.924 \\

\hline \hline 
\end{tabular}
\label{tab:perf-marco-psg}
\end{table*}

\begin{table*}[t]
\caption{MSMARCO document collection results. Results not applicable are denoted `--' and no available `n.a.'.}
\centering
\renewcommand{\arraystretch}{0.9}
\begin{tabular}{ l || c | c c | c c}
\hline \hline
 & \multicolumn{4}{c}{MS MARCO Document Ranking}\\ 
\hline
 & Dev Rerank & \multicolumn{2}{c|}{Dev Retrieval} & \multicolumn{2}{c}{ DL2019 Retrieval} \\ 
 
Model & MRR@10 & MRR@10 & Recall@1K & NDCG@10 & MRR@1K  \\  
\hline
\textbf{Lexical Retriever} & & & & & \\
BM25  & -- &  0.230	& 0.886 & 0.519 &  0.805  \\
DeepCT & -- & 0.320 & 0.942 & 0.544 & 0.891 \\
DocT5Query & -- & 0.288 & 0.926 & 0.597 & 0.837 \\
BM25+BERT reranker  & 0.383 &  --	& -- & -- &  --  \\

\hline
\textbf{Dense Retriever} & & & & & \\
Dense (BM25 neg) & n.a. & 0.299 & 0.928 & 0.600 & n.a. \\
Dense (rand + BM25 neg) & n.a. & 0.311 & 0.952 & 0.576 & n.a. \\
Dense (our train) & 0.358 & 0.340 & 0.883 & 0.546 & 0.785  \\
\hline
COIL-tok & 0.381 & 0.385 & 0.952 & 0.626 & 0.921 \\
COIL-full & 0.388 & 0.397 & 0.962 & 0.636 & 0.913 \\
\hline \hline 
\end{tabular}
\label{tab:perf-marco-doc}
\end{table*}

\begin{table*}[t]
\caption{Performance and latency of COIL systems with different representation dimensions. Results not applicable are denoted `--' and no available `n.a.'. Here $n_c$ denotes COIL CLS dimension and $n_t$ token vector dimension. *: ColBERT use approximate search and quantization. We exclude I/O time from measurements.}
\centering
\renewcommand{\arraystretch}{0.9}
\begin{tabular}{ c  c | c c | c c | c c }
\hline\hline
&  & \multicolumn{2}{c|}{Dev Retrieval} & \multicolumn{2}{c|}{ DL2019 Retrieval} & \multicolumn{2}{c}{Latency/ms}  \\ 
 
\multicolumn{2}{l|}{Model} & MRR@10 & Recall@1K & NDCG@10 & MRR & CPU & GPU  \\  
\hline
\multicolumn{2}{l|}{BM25} & 0.184 & 0.853 & 0.506 & 0.825 & 36 & n.a. \\
\multicolumn{2}{l|}{Dense} & 0.304 & 0.932 & 0.635 & 0.898 & 293 & 32 \\
\multicolumn{2}{l|}{ColBERT} & 0.360 & 0.968 & n.a. & n.a. & 458* & -- \\
\hline
\multicolumn{2}{l|}{COIL} & & & &\\
$n_c$ & $n_t$ & & & & &\\
768 & 32 & 0.355 & 0.963 & 0.704 & 0.924 & 380 & 41 \\
128 & 32 &0.350 & 0.953 & 0.692 & 0.956 & 125 & 23\\
128 & 8 & 0.347 & 0.956 & 0.694 & 0.977 & 113 & 21 \\

0 & 32 & 0.341 & 0.949 & 0.660 & 0.915 & 67 & 18 \\
0 & 8 & 0.336 & 0.940 & 0.678 & 0.953 & 55 & 16 \\

\hline \hline 
\end{tabular}
\label{tab:dimension}
\end{table*}

\paragraph{Compared Systems}
Baselines include 1) traditional exact match system BM25, 2) deep LM augmented BM25 systems DeepCT~\cite{DeepCT} and DocT5Query~\cite{docTTTTTquery},  3) dense retrievers, and 4)  soft all-to-all retriever ColBERT.  For DeepCT and DocT5Query, we use the rankings provided by the authors. For dense retrievers, we report two dense retrievers trained with BM25 negatives or with mixed BM25 and random negatives, published in \citet{ANCE}.  However since these systems use a robust version of BERT, RoBERTa~\cite{RoBERTa} as the LM and train document retriever also on MSMARCO passage set, we in addition reproduce a third dense retriever, that uses the exact same training setup as COIL. All dense retrievers use 768 dimension embedding. For ColBERT, we report its published results (available only on passage collection). BERT reranker is added in the rerank task.

We include 2 COIL systems: 1) COIL-tok, the exact token match only system, and 2) COLL-full, the model with both token match and CLS match. 

\paragraph{Implementation}
We build our models with Pytorch~\cite{pytorch} based on huggingface transformers~\cite{hf-transformers}. COIL's LM is based on BERT's base variant. COIL systems use token dimension $n_t=32$ and COIL-full use CLS dimension $n_c=768$ as default, leading to 110M parameters. We add a Layer Normalization to CLS vector when useful.
All models are trained for 5 epochs with AdamW optimizer, a learning rate of 3e-6, 0.1 warm-up ratio, and linear learning rate decay, which takes around 12 hours. Hard negatives are sampled from top 1000 BM25 results. Each query uses 1 positive and 7 hard negatives; each  batch uses 8 queries on MSMARCO passage  and  4 on MSMARCO document. Documents are truncated to the first 512 tokens to fit in BERT.  We conduct validation on randomly selected 512 queries from corresponding train set. Latency numbers are measured on dual Xeon E5-2630 v3 for CPU and RTX 2080 ti for GPU. 
We implement COIL's inverted lists as matrices as described in \autoref{sec:retrive-and-index}, using NumPy~\cite{numpy} on CPU and Pytorch on GPU. We perform a) a set of matrix products to compute token similarities over contextualized inverted lists, b) scatter to map token scores back to documents, 
and c) sort to rank the documents. Illustration can be found in the appendix, \autoref{fig:index-search}. 


\section{Results}
\label{sec:results}
This section studies the effectiveness of COIL and how vector dimension in COIL affects the effectiveness-efficiency tradeoff. We also provide qualitative analysis on contextualized exact match.

\subsection{Main Results}

 \autoref{tab:perf-marco-psg} reports various systems' performance on the MARCO passage collection. COIL-tok exact lexical match only system significantly outperforms all previous lexical retrieval systems.  With contextualized term similarities, COIL-tok achieves a MRR of 0.34 compared to BM25's MRR 0.18. DeepCT and DocT5Query, which also use deep LMs like BERT and T5, are able to break the limit of heuristic term frequencies but are still limited by semantic mismatch issues. We see COIL-tok outperforms both systems by a large margin. 

COIL-tok also ranks top of the candidate list better than dense retrieves. It prevails in MRR and NDCG while performs on par in recall with the best dense system, indicating that COIL's token level interaction can improve precision.
With the CLS matching added, COIL-full gains the ability to handle mismatched vocabulary and enjoys another performance leap, outperforming all dense retrievers.

COIL-full achieves a very narrow performance gap to ColBERT. Recall that ColBERT computes all-to-all soft matches between all token pairs. For retrieval, it needs to consider for each query token \emph{all} mentions of \emph{all} tokens in the collection (MSMARCO passage collection has around 500M token mentions). COIL-full is able to capture matching patterns as effectively with exact match signals from only query tokens' mentions and a single CLS matching to bridge the vocabulary gap.

We observe a similar pattern in the rerank task. COIL-tok is already able to outperform dense retriever and COIL-full further adds up to performance with CLS matching, being on-par with ColBERT. 
Meanwhile, previous BERT rerankers have little performance advantage over COIL \footnote{Close performance between COIL and BERT rerankers is partially due to the bottleneck of BM25 candidates.}. In practice, we found BERT rerankers to be much more expensive, requiring over 2700 ms for reranking compared to around 10ms in the case of COIL.

\autoref{tab:perf-marco-doc} reports the results on MSMARCO document collection. In general, we observe a similar pattern as with the passage case. COIL systems significantly outperform both lexical and dense systems in MRR and NDCG and retain a small advantage measured in recall. The results suggest that 
COIL can be applicable to longer documents with a consistent advantage in effectiveness. 

The results indicate exact lexical match mechanism can be greatly improved with the introduction of contextualized representation in COIL. COIL's token-level match also yields better fine-grained signals than dense retriever's global match signal. COIL-full further combines the lexical signals with dense CLS match, forming a system that can deal with both vocabulary and semantic mismatch, being as effective as all-to-all system.


\begin{table*}[t]
\caption{Sample query document pairs with similarity scores produced by COIL. Tokens in examination are colored blue. Numbers in brackets are query-document vector similarities computed with vectors generated by COIL.}
\adjustbox{max width=0.98\linewidth}{
\begin{tabular}{l| l | c}
\hline\hline
Query Token  & \multicolumn{1}{c|}{COIL Contextualized Exact Match Score} & Relevance\\ 
\hline 
\multirow{2}{*}{what is a \textbf{\textcolor{blue}{cabinet}}  in govt} 
& \begin{tabular}[l]{p{0.7\textwidth}} \textbf{\textcolor{blue}{\textcolor{blue}{Cabinet  {[}16.28{]} }}} (government) A \textbf{\textcolor{blue}{cabinet {[}16.75{]}}} is a body of high-ranking state officials, typically consisting of the top leaders of the .... \end{tabular} & +\\
\cline{2-3} 
&  \begin{tabular}[l]{p{0.7\textwidth}} \textbf{\textcolor{blue}{Cabinet {[}7.23{]}}} is 20x60 and top is 28x72. .... I had a 2cm granite countertop installed with a 10 inch overhang on one side and a 14 inch.... \end{tabular} & -\\
\hline 
\multirow{2}{*}{what is priority \textbf{\textcolor{blue}{pass}}  } 
& \begin{tabular}[l]{p{0.7\textwidth}} Priority \textbf{\textcolor{blue}{Pass {[}11.61{]}}} is an independent airport lounge access program. A membership provides you with access to their network of over 700 ....\end{tabular} & +\\
\cline{2-3}
& \begin{tabular}[l]{p{0.7\textwidth}} Snoqualmie \textbf{\textcolor{blue}{Pass {[}7.98{]}}} is a mountain \textbf{\textcolor{blue}{pass {[}6.83{]}}} that carries Interstate 90 through the Cascade Range in the U.S. State of Washington....  \end{tabular} & -\\
\hline 
\multirow{2}{*}{what \textbf{\textcolor{blue}{is}} njstart }
& \begin{tabular}[l]{p{0.7\textwidth}} NJSTART \textbf{\textcolor{blue}{is {[}1.25{]}}} a self-service online platform that allows vendors to manage forms, certifications, submit proposals, access training .... \end{tabular} & +\\
\cline{2-3}
& \begin{tabular}[l]{p{0.7\textwidth}} Contract awardees will receive their Blanket P.O. once it \textbf{\textcolor{blue}{is {[}-0.10{]}}} converted, and details regarding that process will also be sent...\end{tabular} & -\\
\hline\hline   
\end{tabular}
}
\label{tab:analysis}
\end{table*}

\subsection{Analysis of Dimensionality}
The second experiment tests how varying COIL's token dimension $n_t$ and CLS dimension $n_c$ affect model effectiveness and efficiency. We record retrieval performance and latency on MARCO passage collection in \autoref{tab:dimension}.

In COIL-full systems, reducing CLS dimension from 768 to 128 leads to a small drop in performance on the Dev set, indicating that a full 768 dimension may not be necessary for COIL. Keeping CLS dimension at 128, systems with token dimension 32 and 8 have very small performance difference, suggesting that token-specific semantic consumes much fewer dimensions. Similar pattern in $n_t$ is also observed in COIL-tok~($n_c=0$). 

On the DL2019 queries, we observe that reducing dimension actually achieves better MRR. We believe this is due to a regulatory effect, as the test queries were labeled differently from the MSMARCO train/dev queries~\cite{craswell2020overview}.

We also record CPU and GPU search latency in \autoref{tab:dimension}.  
Lowering COIL-full's CLS dimension from 768 to 128 gives a big speedup, making COIL faster than DPR system. Further dropping token dimensions provide some extra speedup. The COIL-tok systems run faster than COIL-full, with a latency of the same order of magnitude as the traditional BM25 system. Importantly, lower dimension COIL systems still retain a performance advantage over dense systems while being much faster.
We include ColBERT's latency reported in the original paper, which was optimized by approximate search and quantization. All COIL systems have lower latency than ColBERT even though our current implementation does not use those optimization techniques. We however note that approximate search and quantization are applicable to COIL, and 
 leave the study of speeding up COIL to future work.

\subsection{Case Study}


COIL differs from all previous embedding-based models in that it does not use a single unified embedding space. Instead, for a specific token, COIL learns an embedding space 
to encode and measure the semantic similarity of the token in different contexts. 
In this section, we show examples where COIL differentiates different senses of a word under different contexts.
In \autoref{tab:analysis}, we show how the token similarity scores differ across contexts in relevant and irrelevant query document pairs. 

The first query looks for ``cabinet'' in the context of ``govt'' (abbreviation for ``government''). The two documents both include query token "cabinet" but of a different concept. The first one refers to the government cabinet and the second to a case or cupboard. COIL  manages to match ``cabinet'' in the query to ``cabinet''  in the first document with a much higher score. In the second query, "pass" in both documents refer to the concept of permission. However, through contextualization, COIL captures the variation of the same concept and assigns a higher score to ``pass'' in the first document.  

Stop words like ``it'', ``a'', and ``the'' are commonly removed in classic exact match IR  systems as they are not informative on their own. In the third query, on the other hand, we observe that COIL is able to differentiate ``is''  in an explanatory sentence  and ``is'' in a passive form, assigning the first higher score to match query context.

All examples here show that COIL can go beyond matching token surface form and introduce rich context information to estimate matching. Differences in similarity scores across mentions under different contexts demonstrate how COIL systems gain strength over lexical systems.
\section{Conclusion and Future Work}
Exact lexical match systems have been widely used for decades in classical IR systems and prove to be effective and efficient. 
In this paper, we point out a critical problem, semantic mismatch, that generally limits all IR systems based on surface token for matching. 
To fix semantic mismatch, we introduce contextualized exact match to differentiate the same token in different contexts, providing  effective semantic-aware token match signals. 
We further propose contextualized inverted list~(COIL) search index which swaps token statistics in inverted lists with contextualized vector representations to perform effective search. 

On two large-scale ad hoc retrieval benchmarks, we find COIL substantially improves lexical retrieval and outperforms state-of-the-art dense retrieval systems. 
These results indicate large headroom of the simple-but-efficient exact lexical match scheme.   When the introduction of contextualization handles the issue of semantic mismatch, exact match system gains the capability of modeling complicated matching patterns that were not captured by classical systems. 

Vocabulary mismatch in COIL can also be largely mitigated with a high-level CLS vector matching. The full system performs on par with more expensive and complex all-to-all match retrievers.  
The success of the full system also shows that dense retrieval and COIL's exact token matching give complementary effects, with COIL making up dense system's lost token level matching signals and dense solving the vocabulary mismatch probably for COIL.


With our COIL systems showing viable search latency, we believe this paper makes a solid step towards building next-generation index that stores semantics. At the intersection of lexical and neural systems, 
efficient algorithms proposed for both can push COIL towards real-world systems. 
\newpage
\bibliography{anthology,custom}
 \bibliographystyle{acl_natbib}

\clearpage
\appendix
\onecolumn
\section{Appendix}
\label{sec:appendix}
\subsection{Index Building Illustration}
The following figure demonstrates how the document "apple pie baked ..." is indexed by COIL. The document is first processed by a fine-tuned deep LM to produce for each token a contextualized vector. The vectors of each term "apple" and "juice" are collected to the corresponding inverted list index along with the document id for lookup.
\begin{figure*}[h]
    \centering
    \includegraphics[width=0.8\textwidth]{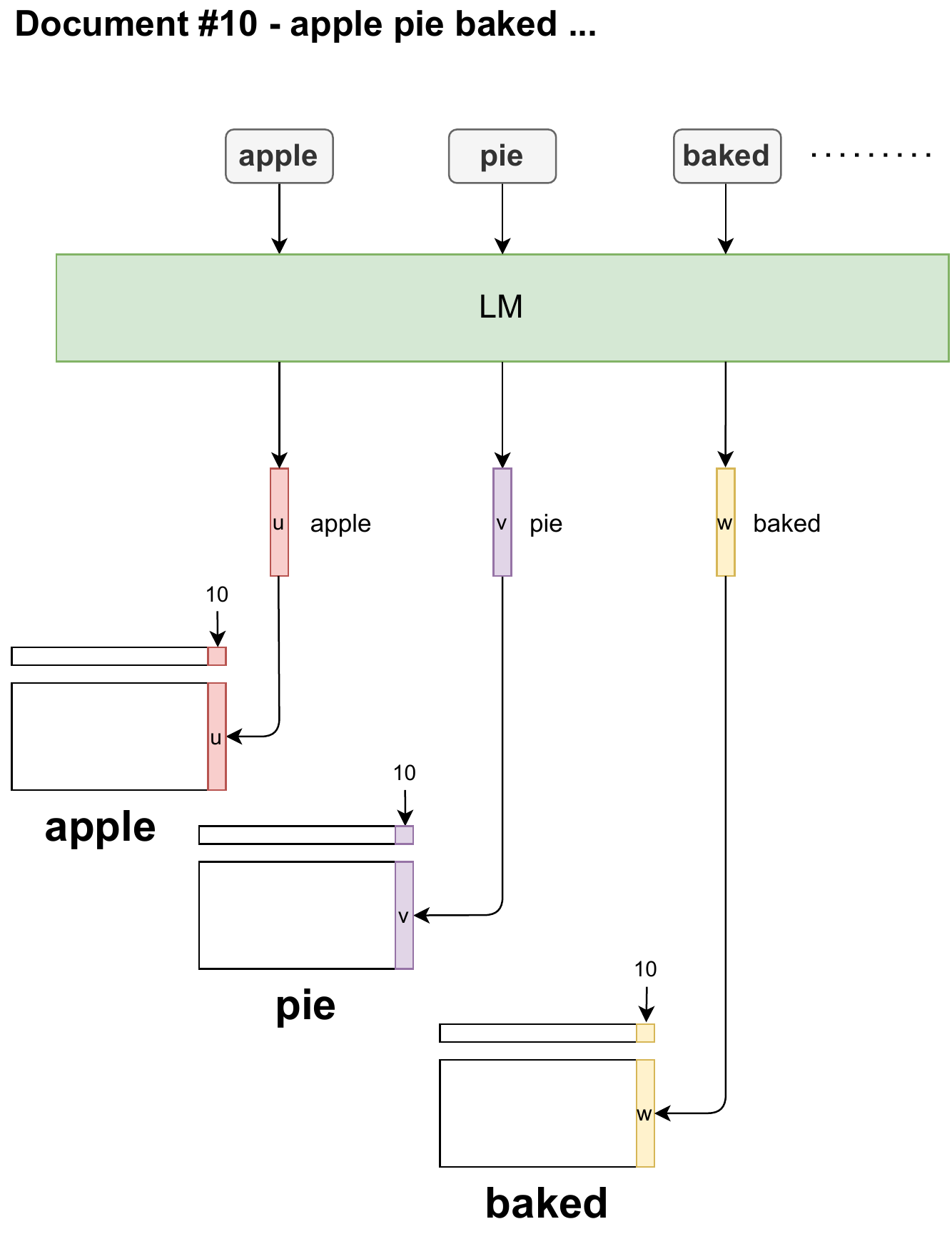}
    \caption{COIL Index Building of document "apple pie baked..."}
    \label{fig:index-build}
\end{figure*}

\newpage
\subsection{Search Illustration}
The following figure demonstrates how the query "apple juice" is processed by COIL. Contextualized vectors of each term "apple" and "juice" go to the corresponding inverted list index consisting of a lookup id array and a matrix stacked from document term vectors. For each index, a \emph{matrix vector product} is run to produce an array of scores. Afterwards a \emph{max-scatter} of scores followed by a \emph{sort} produces the final ranking. Note for each index, we show only operations for a subset of vectors~(3 vectors) in the index matrix.
\begin{figure*}[h]
    \centering
    \includegraphics[width=\textwidth]{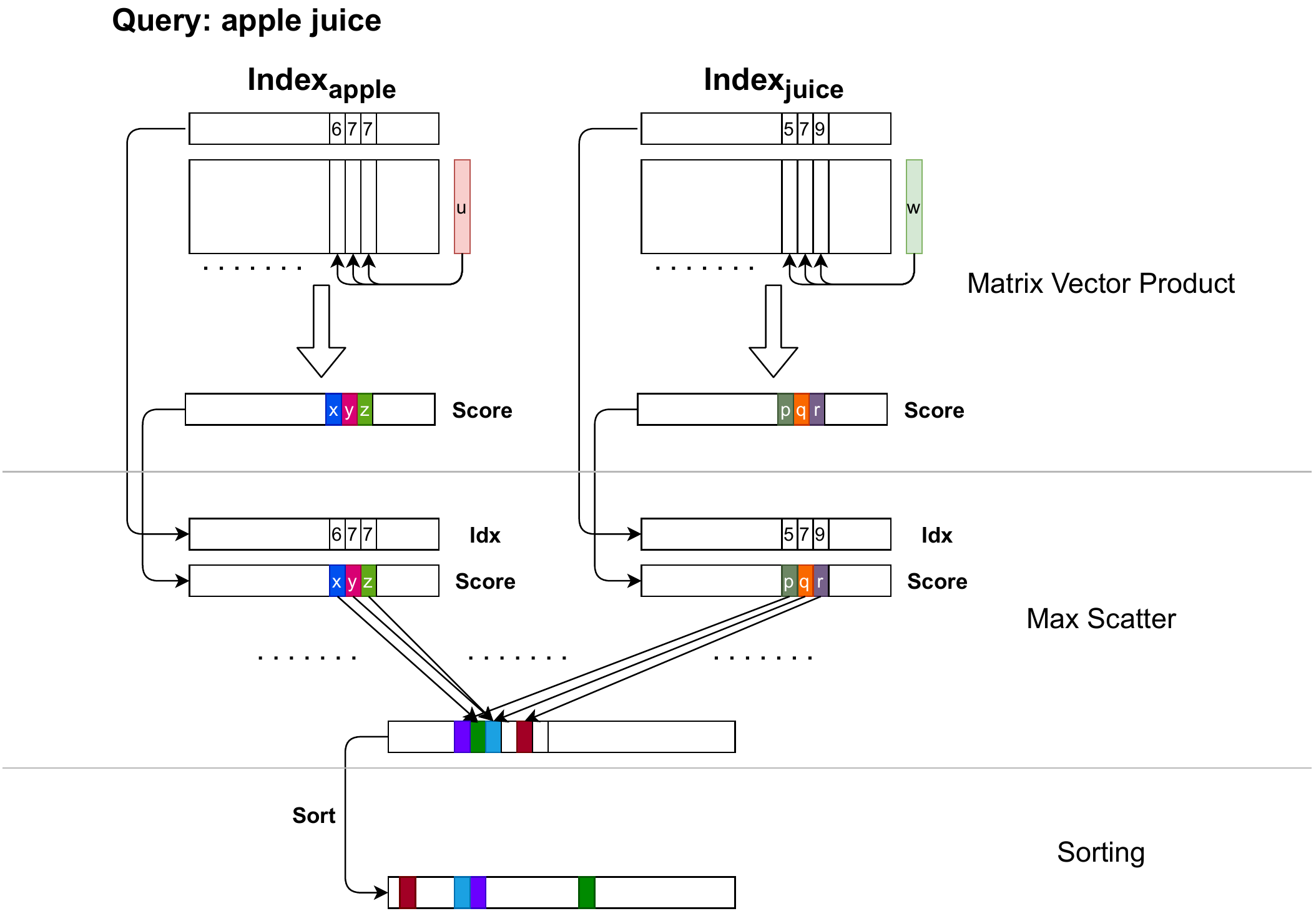}
    \caption{COIL Search of query "apple juice".}
    \label{fig:index-search}
\end{figure*}


\end{document}